\documentclass[aps,prc,notitlepage,nofootinbib,amsmath,amssymb,showpacs,longbibliography,superscriptaddress,twocolumn]{revtex4-1}
\usepackage{graphics}
\usepackage{graphicx}
\usepackage{color}
\usepackage{hyperref}
\usepackage{bm}
\usepackage{calligra}
\usepackage[T1]{fontenc}
\usepackage{amsmath}
\usepackage{amssymb}
\usepackage{amsfonts}
\usepackage{mathrsfs}  

\usepackage{dsfont}% for the identity simbol

\usepackage[utf8]{inputenc}

\usepackage{lipsum}

\usepackage{textcomp}
\usepackage{color}

\begin{document}
%%%%%%%%%%%%%%%%%%%%%%%%%%%%% BIBITEX %%%%%%%%%%%%%%%%%%%%%%%%%%%%%%%%%%%%

\title{Measuring Momentum-Dependent Flow Fluctuations in Heavy-Ion Collisions}

\author{M.~Hippert}
\email{hippert@ifi.unicamp.br}
\affiliation{Instituto de F\'{\i}sica, Universidade de  S\~{a}o Paulo,  Rua  do  Mat\~{a}o, 1371,  Butant\~{a},  05508-090,  S\~{a}o  Paulo,  Brazil}
\affiliation{Instituto de F\'isica Gleb Wataghin, Universidade Estadual de Campinas, Rua S\'ergio Buarque de Holanda 777, 13083-859 S\~ao Paulo, Brazil}

\author{D.D.~Chinellato}
\email{daviddc@g.unicamp.br}
\affiliation{Instituto de F\'isica Gleb Wataghin, Universidade Estadual de Campinas, Rua S\'ergio Buarque de Holanda 777, 13083-859 S\~ao Paulo, Brazil}

\author{M.~Luzum}
\email{mluzum@usp.br}
\affiliation{Instituto de F\'{\i}sica, Universidade de  S\~{a}o Paulo,  Rua  do  Mat\~{a}o, 1371,  Butant\~{a},  05508-090,  S\~{a}o  Paulo,  Brazil}

\author{J.~Noronha}
\email{jn0508@illinois.edu}
\affiliation{Instituto de F\'{\i}sica, Universidade de  S\~{a}o Paulo,  Rua  do  Mat\~{a}o, 1371,  Butant\~{a},  05508-090,  S\~{a}o  Paulo,  Brazil}
\affiliation{Department of Physics, University of Illinois, 1110 W. Green St., Urbana IL 61801-3080, USA}

\author{T.~Nunes da Silva}
\email{t.j.nunes@ufsc.br}
\affiliation{Departamento de  F\'{\i}sica - Centro de Ci\^encias  F\'{\i}sicas e Matem\'aticas, Universidade Federal de Santa Catarina, Campus Universit\'ario Reitor Jo\~ao David Ferreira Lima, Florian\'opolis 88040-900, Brazil}

\author{J.~Takahashi}
\email{jun@ifi.unicamp.br}
\affiliation{Instituto de F\'isica Gleb Wataghin, Universidade Estadual de Campinas, Rua S\'ergio Buarque de Holanda 777, 13083-859 S\~ao Paulo, Brazil}

\date{\today}

\begin{abstract} 
In heavy-ion collisions, momentum-dependent pair correlations can be characterized by a principal component analysis (PCA), 
in which subleading modes are expected to reveal new information on flow fluctuations. However, we find that, as currently 
measured, these modes can be dominated by multiplicity fluctuations, which serve as  an unwanted background. Here, we 
propose new PCA observables that are  robust against multiplicity fluctuations and isolate novel sources of flow 
fluctuations, thus being suited to provide fresh insight into the initial stages of the system at small length scales.
\end{abstract}

\maketitle

%%%%%%%%%%%%%%%%%%%%%%%%%%%%%%%%%%%%%%%%%%%%%%%%%%%%%%%%%%%%%%%%%%%%%%%%%%%%%%

\section{Introduction}

According to the standard picture of relativistic heavy-ion collisions, 
observed momentum anisotropies arise from the hydrodynamic response of the quark-gluon plasma (QGP) 
to the geometry of the initial state \cite{Ollitrault:1992bk,Teaney:2010vd,Gardim:2011xv}. 
As the initial conditions fluctuate event by event, momentum-dependent azimuthal correlations develop 
among particles in their final state.  
The detailed momentum dependence of two-particle correlations is, so far, the only available probe  
of the granularity of initial-state fluctuations, to which other more standard observables are insensitive \cite{Gardim:2017ruc,Kozlov:2014fqa,Noronha-Hostler:2015coa}. Therefore, a thorough study of two-particle correlations and their momentum dependence is needed to resolve the relevant sub-nucleonic length scales present in the initial state of heavy-ion collisions. 

In  Ref.~\cite{Gardim:2012im}, diagonal and off-diagonal momentum-dependent correlations were investigated with the introduction of 
the factorization breaking ratio $r_n( p_{T}^a, p_T^b)$, which measures the correlation of the anisotropic flow between two different transverse momenta 
\cite{Gardim:2012im,Heinz:2013bua,Kozlov:2014fqa,Gardim:2017ruc,Kozlov:2014hya,Shen:2015qta,Zhao:2017yhj,Bozek:2018nne}.  
In Ref.~\cite{Bhalerao:2014mua}, 
the principal component analysis (PCA) of flow fluctuations was proposed to characterize 
the same correlations in a more concise and physically transparent way \cite{Bhalerao:2014mua,Mazeliauskas:2015vea,Mazeliauskas:2015efa,Cirkovic:2016kxt,Bozek:2017thv}. 
Measurements of the factorization breaking ratio and the PCA of event-by-event fluctuations were presented in 
Refs.~\cite{Sirunyan:2017gyb,Khachatryan:2015oea,CMS:2013bza,Acharya:2017ino}.

Principal component analysis is a statistical method used to find linearly uncorrelated combinations of correlated variables  
through the spectral decomposition of the covariance matrix \cite{doi:10.1002/0470013192.bsa501}.  
These combinations, referred to as principal components, are ordered according to their variances, which are eigenvalues of the covariance matrix. 
Hence, this method not only sorts out independent fluctuations but also arranges them according to importance, while requiring no model or assumptions whatsoever. 
In the case of anisotropic flow, the leading PCA component is related to more traditional measures of the flow harmonics, 
while the subleading ones are supposed to carry more detailed information on momentum-dependent flow fluctuations \cite{Sirunyan:2017gyb,Bhalerao:2014mua,Cirkovic:2016kxt}. 

In this paper, we discuss a crucial issue with the interpretation of the PCA observables as defined in Ref.~\cite{Bhalerao:2014mua}. 
This issue, related to multiplicity fluctuations,  
leads to unexpected redundancies, which we illustrate using experimental data from the CMS Collaboration \cite{Sirunyan:2017gyb}. In this paper we introduce a new version of the PCA observables that factors out those redundancies and properly reflects momentum-dependent anisotropic flow fluctuations. We believe the redefined observables to be the ultimate tool for investigating two-particle correlations. By isolating new sources of correlation, they could be particularly useful for constraining properties of the QGP and its initial stages.

After hydrodynamic expansion,  
particles at a given transverse momentum $p_T$ and  rapidity $y$ 
 are azimuthally distributed according to a probability distribution, 
\begin{equation}
 E\,\dfrac{d N}{d^3 p} =\dfrac{1}{2\pi} 
 N(p_T,y) \sum_{n=-\infty}^{\infty} {V}_n(p_T,y)\,e^{- i n \varphi}\,,
 \label{eq:FourierYield}
\end{equation}
where $V_n(p_T,y)$ are complex flow vectors and 
$N(p_T,y)$ is the particle density as a function of rapidity and transverse momentum.  
Due to the random orientation of events, $\langle V_n (p_T,y)\rangle = 0$, 
where $\langle \cdots \rangle$ represents an average over events.  
While particles are understood as independent samples of distribution (\ref{eq:FourierYield}), 
the limited number of particles makes it impossible to accurately measure $V_n(p_T,y)$ in each event.
One must instead study its statistical properties in an ensemble of events via multiparticle correlations.

We are interested in fluctuations of the flow anisotropy as a function of momentum, 
given by the harmonics $V_n(p_T,y)$.  
The main features of these fluctuations are captured by the covariance matrix 
\begin{equation}
 V_{n\Delta} (\bm p_1,\bm p_2)\equiv \langle V_n^*(\bm p_1)\, V_n(\bm p_2) \rangle\,,
 \label{eq:def-Vcov}
 \end{equation}
where we denote ${\bm p}\equiv (p_T,y)$. 
Note that flow fluctuations are correlated across different momenta and $V_{n\Delta}(\bm p_1,\bm p_2)$ is 
a nondiagonal matrix. 
In case there is only one source of fluctuations (e.g., fluctuations of the system orientation), $V_{n\Delta}$ has 
only one nonvanishing eigenvalue and factorizes as 
\begin{equation}
 V_{n\Delta} (\bm p_1,\bm p_2)\;\mathrel{\stackrel{\makebox[0pt]{\mbox{\normalfont\tiny fact.}}}{=}}\;\sqrt{\langle |V_n(\bm p_1)|^2\rangle}\, \sqrt{\langle|V_n(\bm p_2)|^2 \rangle}\,.
  \label{eq:factorization}
\end{equation}
However, this factorization is not perfect, indicating more than one relevant eigenvalue \cite{Gardim:2012im,Bhalerao:2014mua}. 
In general, each nonvanishing eigenvalue of a covariance matrix corresponds to a linearly uncorrelated 
fluctuation mode. 
The subdominant fluctuations signaled by the breaking of Eq.~(\ref{eq:factorization}) 
imply that particles of different momenta respond differently to initial-state fluctuations.

In Eq.~(\ref{eq:FourierYield}), the distribution of particles also depends on 
the density of particles $N(\bm p)$. 
Fluctuations of $N(\bm p)$ can be studied similarly with the covariance matrix \cite{Bhalerao:2014mua}
\begin{equation}
  N_{\Delta} (\bm p_1,\bm p_2)\equiv \langle \Delta N(\bm p_1)\, \Delta N(\bm p_2) \rangle\,,
   \label{eq:def-Ncov}
\end{equation}
where $\Delta(\cdots)\equiv (\cdots) - \langle( \cdots) \rangle$. 
This covariance matrix 
contains nontrivial 
information about fluctuations of mean transverse momentum and
their correlation with global multiplicity fluctuations 
\cite{Gardim:2019iah}.

\section{Principal component analysis}

 The principal component analysis of event-by-event fluctuations was introduced in Ref.~\cite{Bhalerao:2014mua}.  
 The idea is to isolate the linearly independent fluctuation modes contributing to $V_{n\Delta}$ 
by computing its eigenvalues and eigenvectors.
  According to the spectral theorem, 
 \begin{equation}
 \begin{split}
  {V}_{n\Delta}(\bm p_a,\bm p_b) &= \sum_{\alpha=1}^\infty \lambda^{(\alpha)}_n\, \psi_n^{(\alpha)}(\bm p_a)\, \psi_n^{(\alpha)}(\bm p_b)\\
 &= \sum_{\alpha=1}^\infty {V}_n^{(\alpha)}(\bm p_a)\, {V}_n^{(\alpha)}(\bm p_b)
  \,,
  \label{eq:VnSpectTheorem}
 \end{split}
 \end{equation}
 where $\lambda^{(\alpha)}_n$  and $ \psi_n^{(\alpha)}(\bm p)$ are the eigenvalues and normalized eigenvectors of $V_{n\Delta}$. 
 Because $\operatorname{Im} [V_n^*(\bm p_1) V_n(\bm p_2)]$  is odd under 
 parity transformations,  
$V_{n\Delta}$ must be real \cite{Gardim:2012im,Bhalerao:2014mua}. 
Moreover, a covariance matrix must have positive eigenvalues. 
  The principal components, or modes, of the flow fluctuations are defined as
\begin{equation}
{V}_n^{(\alpha)}(\bm p)\equiv\sqrt{\lambda^{(\alpha)}_n}\, \psi_n^{(\alpha)}(\bm p)\,.
\label{eq:defvmode}
\end{equation}
Labeling the eigenvalues in descending order, $\lambda^{(\alpha)}_n \geq \lambda^{(\alpha+1)}_n$, an approximation can be found by  truncating the 
sum in Eq.~(\ref{eq:VnSpectTheorem}) at a given cut $\alpha = \alpha_{\textrm{max}}$.

The physical interpretation of the PCA can be clarified by projecting $V_n(\bm p)$ onto the basis defined by $\{V_n^{(\alpha)}(\bm p)\}$:
  \begin{equation}
  {V}_n(\bm p)
  \approx \sum_{\alpha=1}^{\alpha_{\textrm{max}}} \xi^{(\alpha)}_n\,{V}_n^{(\alpha)}(\bm p)\,,
  \label{eq:flowfromPCA}
  \end{equation}
  where $\langle \xi^{(\alpha)}_n\rangle =0$. 
From Eqs.~(\ref{eq:def-Vcov}) and (\ref{eq:VnSpectTheorem}) one finds that 
$\langle \xi^{(\alpha)*}_n\,\xi^{(\beta)}_n\rangle = \delta_{\alpha\beta}$. 
Indeed, the PCA isolates linearly uncorrelated fluctuation modes, with both their magnitudes and momentum dependence characterized by 
${V}_n^{(\alpha)}(\bm p)$. 
Together with Eq.~(\ref{eq:flowfromPCA}), it allows for an event-by-event description of $V_n(\bm p)$, providing unique insight into the momentum dependence of flow fluctuations.

\begin{figure*}
\centering
   \includegraphics[width=\textwidth]{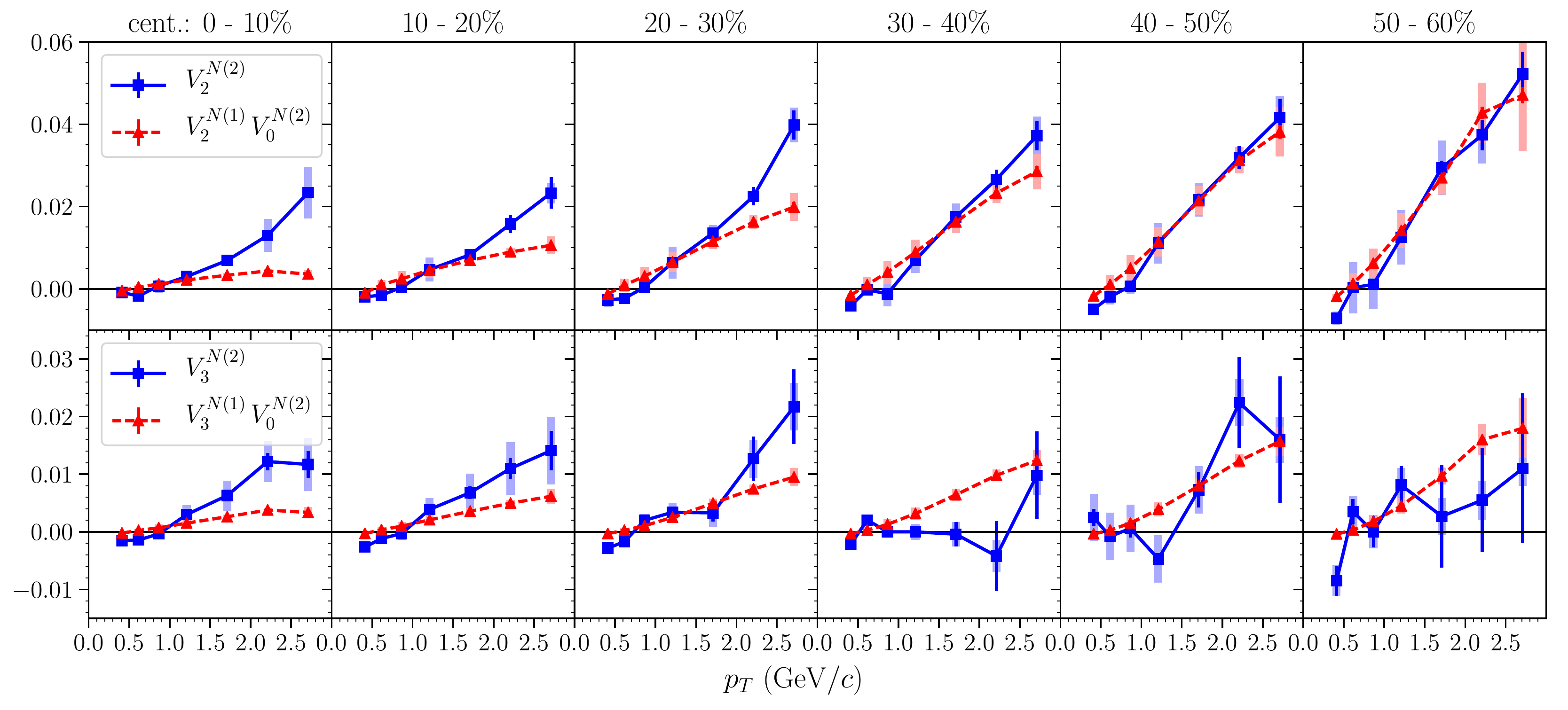}
   \caption{Comparison between the subleading principal component of  $V_{n\Delta}^{N}$ (solid blue curve) and the expected contributions $V_n^{N(1)}\,V_0^{N(2)}$ 
   from multiplicity fluctuations (dashed red curve), using the CMS Collaboration results for $Pb+Pb$ collisions at center-of-mass energy $\sqrt{s_{NN}} = 2.76$ TeV \cite{Sirunyan:2017gyb}. 
   In the second curve we assumed independent variables when propagating uncertainties. 
   The error bars represent statistical uncertainties, while the boxes represent systematic ones. 
}
   \label{fig:PCAmultifluct-CMS}
 \end{figure*}

There is an important subtlety regarding the measurement  $V_{n\Delta}$ and the PCA. 
The direct way to measure the flow covariance matrix is 
\begin{equation}
 {V}_{n\Delta}(\bm p_a,\bm p_b)= \left\langle 
\frac{\sum_{a\neq b} e^{-i n(\phi_a-\phi_b)}}{ N_{\textrm{pairs}}(\bm p_a, \bm p_b)}
\right\rangle\,,
 \label{eq:realdef-Vcov}
\end{equation}
where we sum over the $N_{\textrm{pairs}}$ pairs of particles $a\neq b$ that can be formed between two bins, 
centered around $\bm p_a$ for particle $a$ and $\bm p_b$ for particle $b$.   
In the hydrodynamic picture, particles are independently emitted from the fluid and Eq.~(\ref{eq:def-Vcov}) is retrieved.
However, the principal component analysis of Refs.~\cite{Bhalerao:2014mua,Sirunyan:2017gyb} considered instead the covariance matrix 
\begin{equation}
 \begin{split}
  {V}_{n\Delta}^{N}(\bm p_a,\bm p_b) &\equiv \dfrac{1}{(2\pi\Delta p_T\Delta y)^2} \left\langle \sum_{a\neq b} e^{-i n(\phi_a-\phi_b)}\right\rangle\\
  &\mathrel{\stackrel{\makebox[0pt]{\mbox{\normalfont\tiny hydro}}}{=}}\;\langle N(\bm p_a)\, {V}_n^*(\bm p_a) \,N(\bm p_b)\, {V}_{n}(\bm p_b)\rangle \,.
  \label{eq:flowsum}
 \end{split}
\end{equation}
 Here, $N(\bm p_a)=N_a$ is the multiplicity in bin $a$ normalized by $2\pi\Delta p_T^a\Delta y^a$, 
 where  $\Delta p_T^a$ and $\Delta y^a$ are the bin widths in transverse momentum and pseudorapidity, respectively.\footnote{
 The  normalization of $N$ with $(2\pi\Delta p_T\Delta y)^{-1}$ is chosen for compatibility with Ref.~\cite{Bhalerao:2014mua}, but is not relevant for $V_{n\Delta}$ and $V_{n\Delta}^R$ below. }   
The extra factors of particle number are only compensated at the end of the analysis, by dividing the 
resulting modes by $\langle N(\bm p)\rangle$. 
Thus, one obtains a quantity to be compared to the usual ``per particle'' flow: 
\begin{equation}
{V}_n^{N(\alpha)}(\bm p)\equiv
\sqrt{\lambda^{N(\alpha)}_n}\, \psi_n^{N (\alpha)}(\bm p)/\langle N(\bm p) \rangle\,,
\label{eq:defvmodeN}
\end{equation}
where $\lambda^{N(\alpha)}_n$ and  $\psi_n^{N (\alpha)}(\bm p)$ are the corresponding eigenvalues and eigenvectors.  

An important advantage of Eq.~(\ref{eq:flowsum}) over Eq.~(\ref{eq:realdef-Vcov}) is that it gives more weight 
to events with a larger number of pairs, where the relative uncertainty in the flow vector is smaller. 
However, there are very important differences in the diagonalization of  ${V}_{n\Delta}$  and $ {V}_{n\Delta}^{N}$,  
especially due to the fact that $\langle N_a\, N_b \rangle \neq \langle N_a\rangle\,\langle N_b \rangle$.   
These particle number fluctuations can be measured directly by performing the PCA on the matrix $N_{\Delta}(\bm p_a,\bm p_b) =  {V}_{0\Delta}^{N}(\bm p_a,\bm p_b)$  
and, thus, they represent an unwanted background for $n\neq 0$ analyses.

\section{Multiplicity fluctuations}

The PCA of $V_{0\Delta}^N$, corresponding to particle number fluctuations, was 
investigated in Refs.~\cite{Bhalerao:2014mua,Sirunyan:2017gyb,Gardim:2019iah}. 
While the leading mode is nearly constant in transverse momentum,  the subleading one 
  displays a significant $p_T$ dependence. 
We shall show that momentum-dependent multiplicity fluctuations 
have startling consequences for the $n\neq 0$ PCA of Refs.~\cite{Bhalerao:2014mua,Sirunyan:2017gyb}. 

Figure~\ref{fig:PCAmultifluct-CMS} displays PCA data from the CMS Collaboration, obtained using Eqs.~(\ref{eq:flowsum}) and (\ref{eq:defvmodeN}) \cite{Sirunyan:2017gyb}.  
The solid curves are the measured subleading PCA modes of the elliptic and triangular flow, $V_{2}^{N(2)}$ and $V_{3}^{N(2)}$, while the dashed curves represent the combination $V_n^{N(1)}\,V_0^{N(2)}$, 
the product of the leading mode of the $n$th harmonic  times the subleading mode of particle number fluctuations. 
A striking proximity between the two curves is verified and deserves to be investigated, especially for noncentral collisions and $n=2$. 
The relationship between $ V_n^{N(2)}$ and $V_n^{N(1)}\,V_0^{N(2)}$ in hydrodynamic simulations was first studied 
in Ref.~\cite{Mazeliauskas:2015efa}. In that paper, 
the subleading mode of elliptic flow, $V_2^{N(2)}$, in peripheral collisions was found to be dominated by the nonlinear mixing between leading elliptic flow and subleading radial flow 
fluctuations [i.e., fluctuations of $V_0(p_T)$, captured by $V_0^{N(2)}(p_T)$]. 
Figure~\ref{fig:PCAmultifluct-CMS} displays similar findings, albeit we draw very different conclusions from them.   

To investigate the coincidence in Fig.~\ref{fig:PCAmultifluct-CMS}, let us now 
estimate the contributions of particle-number fluctuations to the flow covariance matrix $V_{n\Delta}^N$. 
Applying Eqs.~(\ref{eq:flowfromPCA}) and (\ref{eq:defvmodeN}) to the PCA of $N_\Delta=V_{0\Delta}^N$, we can estimate the event-by-event differential multiplicity: 
\begin{equation}
 N  \approx \langle N \rangle\,\left( 1 + V_0^{N (1)}\,\xi^{N (1)} + V_0^{N (2)}\,\xi^{N (2)} \right)\,,
 \label{eq:NfromPCA}
 \end{equation}
where, for brevity, we omit the momentum dependence. 
Substituting Eq.~(\ref{eq:NfromPCA}) in Eq.~(\ref{eq:flowsum}) and isolating contributions from $V_n^{(2)}$, we find 
\begin{multline}
 V_{n\Delta}^N(\bm p_1,\bm p_2) \approx \langle N(\bm p_1) \rangle \langle N(\bm p_2) \rangle \left(q^{(1)}(\bm p_1)\,q^{(1)}(\bm p_2) \right. \\
 + \left. q^{(2)}(\bm p_1)\,q^{(2)}(\bm p_2)\right) + \mathcal{O}(V_n^{(2)})\,,
\end{multline}
with
\begin{equation}
 \begin{split}
   q^{(1)}(\bm p) &= \sqrt{1 + \left(V_0^{N (1)}(\bm p_0)\right)^2}\,V_n^{(1)}(\bm p)\,,\\
  q^{(2)}(\bm p) &= V_0^{N (2)}(\bm p)\,V_n^{(1)}(\bm p)\,, 
 \end{split}
 \label{eq:defqs}
\end{equation}
where we assume that fluctuations of $N$ and $V_n$ are independent and that $V_0^{N (1)}(\bm p)$ is constant. 
Because $V_0^{N(1)}$ is nearly constant,  $V_0^{N(1)} V_n^{N(1)}$ is approximately parallel to $V_n^{N(1)}$ 
and contributes to this mode.
On the other hand, this is clearly not the case for $V_0^{N(2)} V_n^{N(1)}$,  
which should contribute to the subleading component. 
Thus, in a first approximation, it is reasonable to expect 
\begin{equation}
\begin{split}
  V_n^{N (1)}(\bm p) \approx q^{(1)}(\bm p) + \mathcal{O}(V_n^{(2)}) \,,\\
   V_n^{N (2)}(\bm p) \approx q^{(2)}(\bm p) + \mathcal{O}(V_n^{(2)})\,.
\end{split}
\label{eq:fakemodes}
\end{equation}

\begin{figure*}
  \centering
   \includegraphics[width=\textwidth]{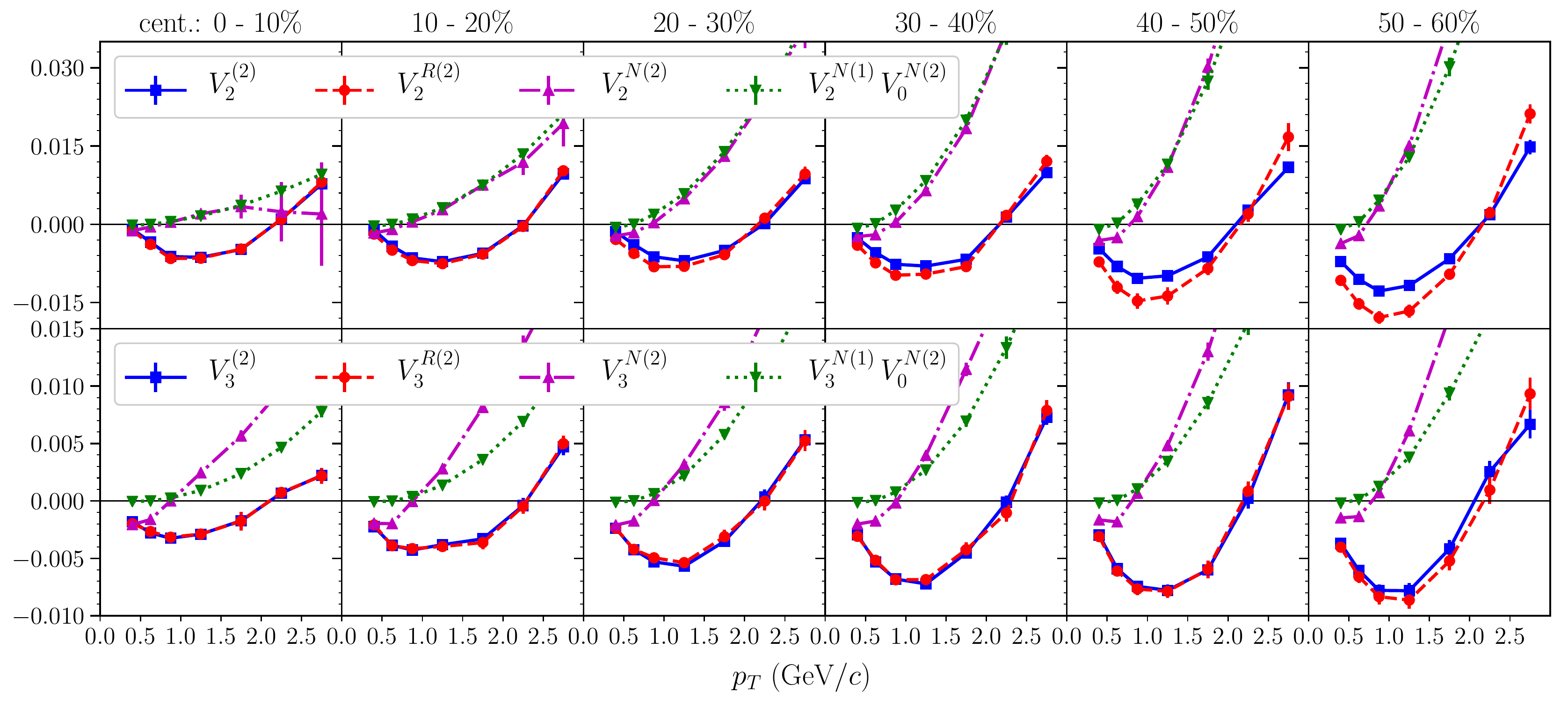}
   \caption{New subleading principal component from $V_{n\Delta}^{R}$ (dashed red curve) and $V_{n\Delta}$ (solid blue curve) in a hybrid hydrodynamic model, for  
   $Pb+Pb$ collisions at center-of-mass energy $\sqrt{s_{NN}} = 2.76$ TeV. 
Error bars represent statistical uncertainties. 
   In general, $V_{n}^{R(2)}$ is seen to match $V_{n}^{(2)}$ to very good approximation. Notice also the relative stability against changes in centrality. 
      For reference, $V_{n}^{N(2)}$ (dash-dotted magenta curve) and the 
   combination $V_n^{N(1)}\,V_0^{N(2)}$ (dotted green curve) are also shown.  
   The behavior observed in Fig.~\ref{fig:PCAmultifluct-CMS} for the CMS data is clearly reproduced by the model. 
   }
   \label{fig:PCA-R-hydro}
 \end{figure*}

Equation~(\ref{eq:fakemodes}) shows that multiplicity fluctuations may severely affect the observables defined in Ref.~\cite{Bhalerao:2014mua}. 
While the correction to the leading PCA mode is small, of at most $\sim 3\%$ for $V_0^{N(1)} \lesssim 0.25$, this is not the case for subleading modes, 
which might be dominated by fluctuations of $N(\bm p)$. 
Taking $V_n^{(2)}\to 0$, we can isolate the contribution of multiplicity fluctuations to  $V^{N(2)}_n$:
\begin{equation}
 V^{N(2)}_n (\bm p) \approx V_0^{N (2)}(\bm p)\,V_n^{N(1)}(\bm p) + \mathcal{O}(V_n^{(2)})\,.
 \label{eq:submultfluct}
\end{equation}
This is precisely the combination shown to approximate the PCA data in Fig.~\ref{fig:PCAmultifluct-CMS}. 
Given the simplicity of the estimate in Eq.~\eqref{eq:submultfluct}, the proximity between the two curves in Fig.~\ref{fig:PCAmultifluct-CMS} is truly striking. 
The very fact that they are comparable in magnitude suggests that multiplicity fluctuations might be responsible for a considerable fraction of $V_n^{N(2)}$. 
In fact, at least for noncentral collisions, $V_2^{N(2)}$ appears to be dominated by these multiplicity fluctuations, rather than fluctuations in the  anisotropic flow itself.

 \section{New set of PCA observables}

The results above are critical to the interpretation of the current PCA of flow harmonic data. 
First, they compromise the interpretation of the subleading PCA modes as revealing entirely new $p_T$-dependent anisotropic flow fluctuations. 
Furthermore, they suggest that these quantities are dominated by redundant information --- that is, information that is more directly obtained from existing measurements of the leading anisotropic mode and the multiplicity PCA modes. 

The flow vector per particle $V_n(\bm p)$ is the quantity most directly related to spatial anisotropy in the initial state, as characterized by geometric eccentricities. 
It is thus desirable to redefine the PCA observables so as to study fluctuations of this quantity alone, instead of the combination $N(\bm p) V_n(\bm p)$  --- removing known contributions from multiplicity fluctuations.  
One possibility is measuring the PCA of $V_{n\Delta}$, as defined in Eqs.~(\ref{eq:VnSpectTheorem}), (\ref{eq:realdef-Vcov}), and (\ref{eq:defvmode}). 
This has the disadvantage of giving the same weight to all events, regardless of the number of particles, which might render $V_{n\Delta}$ less stable. 
However, because statistical uncertainties in our simulations do not accurately represent that from a realistic measurement, we refrain from estimating this effect. 
The PCA of $V_{n\Delta}$ was employed in Ref.~\cite{Bozek:2017thv}.

As an alternative to $V_{n\Delta}^{N}$, we propose the diagonalization of the matrix  
\begin{equation}
 \begin{split}
  {V}_{n\Delta}^{R}(\bm p_a,\bm p_b) &\equiv  
  \left\langle \sum_{a\neq b} e^{-i n(\phi_a-\phi_b)}\right\rangle /\left\langle N_{\textrm{pairs}}(\bm p_a, \bm p_b)\right\rangle \\
  &\mathrel{\stackrel{\makebox[0pt]{\mbox{\normalfont\tiny hydro}}}{=}}\;\dfrac{\left\langle  N(\bm p_a)\, {V}_n^*(\bm p_a)\,N(\bm p_b)\, {V}_n(\bm p_b) \right\rangle}
  {\left\langle  N(\bm p_a)\, N(\bm p_b) \right\rangle}\,,
 \label{eq:pseudonorm-covmat}
 \end{split}
\end{equation}
so that the average is weighted by the number of pairs. 
Equation~(\ref{eq:pseudonorm-covmat}) is the definition of the correlation matrix
that is typically
 considered in measurements of the factorization breaking coefficient 
$r_n(p_{T}^a, p_T^b)$ \cite{Heinz:2013th,CMS:2013bza,Khachatryan:2015oea,Acharya:2017ino}. 
For our purposes, it will provide a good approximation to $V_{n\Delta}$, as will be shown. 
The principal components of $V_{n\Delta}^R$ are given by 
\begin{equation}
{V}_n^{R(\alpha)}(\bm p)\equiv\sqrt{\lambda^{R(\alpha)}_n}\, \psi_n^{R (\alpha)}(\bm p)\,,
\label{eq:defvmodeR}
\end{equation}
where $\lambda^{R(\alpha)}_n$ and  $\psi_n^{R (\alpha)}(\bm p)$ are the corresponding eigenvalues and eigenvectors.

Thus, each pair of particles is given the same weight, but multiplicity fluctuations are canceled by the denominator if they factor out.   That is, if 
\begin{align}
\bigl\langle  N(\bm p_a)\, {V}_n^*(\bm p_a)\,N(\bm p_b)\, {V}_n(\bm p_b) \bigr\rangle& \nonumber\\
\simeq \bigl\langle  N(\bm p_a)N(\bm p_b)\bigr\rangle&\bigl\langle {V}_n^*(\bm p_a) {V}_n(\bm p_b) \bigr\rangle,
\end{align}
then the multiplicity factor cancels and Eq.~\eqref{eq:pseudonorm-covmat} becomes Eq.~\eqref{eq:def-Vcov}.

To test this, we employ a state-of-the-art hybrid  model, in satisfactory agreement with experimental data \cite{NunesdaSilva:2018viu}. 
Our model consists of relativistic viscous hydrodynamics as implemented in {\small MUSIC} \cite{Schenke:2010nt,Schenke:2011bn} and  
 evolution of the hadron gas phase according to UrQMD \cite{Bass:1998ca,Bleicher:1999xi}. 
 Initial conditions were provided by {\small TRENTO} \cite{Moreland:2014oya} and 
parameter values taken from the Bayesian analysis of Ref.~\cite{Bernhard:2018hnz}. 
In Fig.~\ref{fig:PCA-R-hydro}, we compare the new subleading PCA mode ${V}_n^{R(2)}(p_T)$ to the subleading PCA mode of the flow per particle ${V}_n^{(2)}(p_T)$,  obtained from the covariance matrix $V_{n\Delta}$, in Eq.~\eqref{eq:realdef-Vcov}.  
A good agreement is found in general, indicating that  $V_{n\Delta}^R$ can be used in place of  $V_{n\Delta}$. 

Results for the original PCA modes $V_n^{N(2)}$ are also shown in Fig.~\ref{fig:PCA-R-hydro}, together with the combination of Eq.~\eqref{eq:submultfluct}.
It is clear that the new observables behave very differently from the ones in Ref.~\cite{Bhalerao:2014mua}. 
While definition (\ref{eq:pseudonorm-covmat}) does not necessarily remove all effects from multiplicity fluctuations, it does remove the most trivial ones 
from the corresponding two-point function. 
Indeed, $V_{n\Delta}^{R} \approx V_{n\Delta}$ to good approximation unless anisotropic flow and radial flow fluctuations are strongly correlated. 
Figure~\ref{fig:PCA-R-hydro} suggests, however, that this is not the case, because $V_{n}^{R(2)}(p_T)$ closely follows $V_{n}^{(2)}(p_T)$.

\section{Spectral decomposition}

An important subtlety regarding a PCA analysis is that
the spectral decomposition is not unique.  
We can define a general eigenvalue and eigenvector equation as
\begin{equation}
\label{eigen}
\int_\Omega d{\bm p_b} \, W(\bm p_b)\, V_{n\Delta}(\bm p_a, \bm p_b)  V_n^{(\alpha)}(\bm p_b) =\lambda_n^{(\alpha)}  V_n^{(\alpha)}(\bm p_a),
\end{equation}
with an arbitrary weight function $W$ and integration range $\Omega$, both of which will affect the resulting decomposition.  The eigenvectors are then orthogonal with respect to the inner product 
\begin{equation}
\label{innerprod}
 (V_n^{(\alpha)},V_n^{(\beta)}) \equiv \int_\Omega  d{\bm p}\,W(\bm p)\, V_n^{(\alpha)}(\bm p) V_n^{(\beta)}(\bm p)\, = \lambda_n^{(\alpha)}\delta_{\alpha,\beta}.
\end{equation}

In practice, one uses finite-sized bins in momentum space and the integrals become sums over discrete momentum indices, i.e.,
\begin{equation}
\label{eigendisc}
\sum_{b} \Delta{\bm p_b} W(\bm p_b)    V_{n\Delta}(\bm p_a, \bm p_b)  V_n^{(\alpha)}(\bm p_b) = \lambda_n^{(\alpha)}  V_n^{(\alpha)}(\bm p_a).
\end{equation}

A change in the weight $W(\bm p)$ will especially affect the subleading mode, which is related to the 
dominant leading mode by the orthogonality relation $(V_n^{(1)},V_n^{(2)})=0$. 

A calculation can be done without considering the bin width $\Delta{\bm p_a}$ in Eq.~\eqref{eigendisc}. 
This was, in fact, done in Refs.\ \cite{Bhalerao:2014mua,Sirunyan:2017gyb}, such that $W(\bm p)\Delta{\bm p_a} =1$.  
In this case,  the result will change if the bin widths change in a nonuniform way and will depend on the specific choice of binning.  
When using Eq.~\eqref{eigendisc}, on the other hand, the result is stable under any choice of binning, given a fixed choice of $W(\bm p)$.

In Fig.~\ref{fig:PCA-R-hydro}, we employed a weight function $W(\bm p) =1$, 
representing a straightforward spectral decomposition of the covariance matrix of traditional anisotropic flows $V_n$, which therefore also can be  directly connected to the initial state \cite{inprogress}.
This choice results in a dependence on the maximum transverse momentum used in the analysis, which was not present in the original PCA observable \cite{Bhalerao:2014mua}.  
This dependence can be removed, without reintroducing multiplicity fluctuation contamination, with a suitable choice of $W$.  For example, choosing a weight proportional to the (average) particle density,  
$W(\bm p) = \langle N(\bm p) \rangle$, reduces the contribution of particles with large transverse momentum.
We found that the conclusions taken from Fig.~\ref{fig:PCA-R-hydro} hold under this choice as well.

\begin{figure*}
  \centering
   \includegraphics[width=\textwidth]{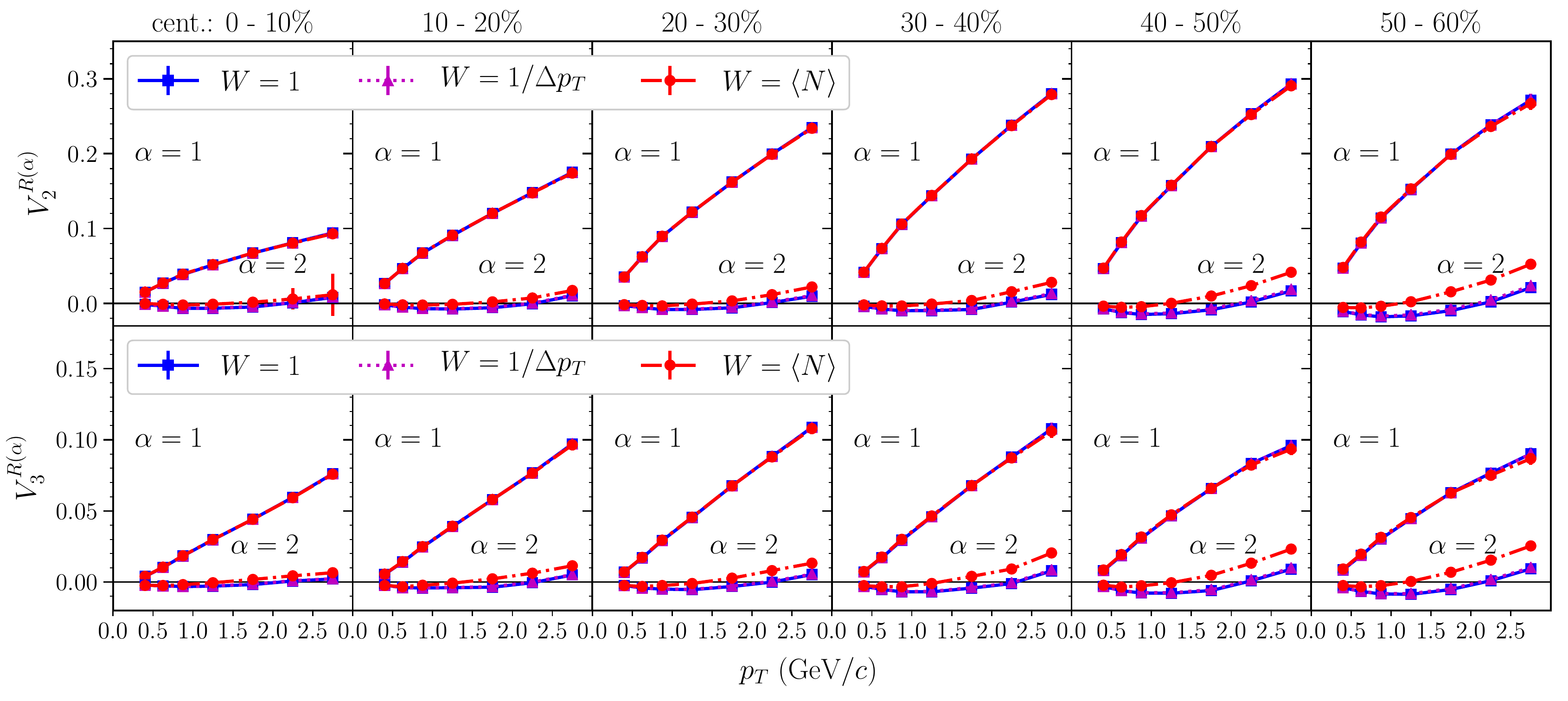}
   \caption{
   Comparison between PCA results  
   with different choices of weight for the spectral decomposition: 
   the straightforward choice, employed in this paper ($W=1$, solid blue curve); 
   the choice adopted in Refs.~\cite{Bhalerao:2014mua} and \cite{Sirunyan:2017gyb} ($W = 1/\Delta p_T$, dashed magenta curve); 
   the alternative choice of using the average charged-particle multiplicity per bin as the weight function (dot-dashed red curve).  
   Both the leading and subleading principal components
   of the matrix $V_{n\Delta}^R$ are displayed for hydrodynamic simulations ({\small TRENTO}+{\small MUSIC}+UrQMD)
   of $Pb+Pb$ collisions at center-of-mass energy $\sqrt{s_{NN}} = 2.76$ TeV, 
   for $n=2$ (upper panels) and $n=3$ (lower panels).
   }
   \label{fig:spectraldec}
 \end{figure*}
 
Figure~\ref{fig:spectraldec} shows the effect of different choices of spectral decomposition. 
In this figure, we can see that, while the effect of the $\Delta p_T$ factor is subtle for our choice of binning, 
employing different weights can have an important effect in the final observables. 

Notice that, in Fig.~\ref{fig:PCA-R-hydro}, $V_n^{N(2)}$ does not seem to converge towards $V_n^{(2)}$ or $V_n^{R(2)}$ at central collisions, where multiplicity 
fluctuations should become less relevant. In fact, in the limit where multiplicity fluctuations vanish, the former should become equivalent to 
the latter, but within a spectral decomposition different from that employed in Fig.~\ref{fig:PCA-R-hydro}. 
This is so because $V_n^{N(2)}$ gives more emphasis to bins of low transverse momentum, where more particles are present, whereas $V_n^{R(2)}$ gives the same weight 
to all momentum bins, unless a weight of $W(p_T) \neq 1$ is employed
As a matter of fact, employing a weight of $W(p_T) = N (p_T)$ in the spectral decomposition for $V_n^{R(2)}$ (see the dot-dashed red curve in Fig.~\ref{fig:spectraldec}) 
already makes results much more similar to the ones for $V_n^{N(2)}$ as seen in Figs.~\ref{fig:PCAmultifluct-CMS} and \ref{fig:PCA-R-hydro}, 
even though the corresponding spectral decompositions are still not equivalent.

\section{Final remarks}

In this paper, we discussed the effect of multiplicity fluctuations in 
the PCA of anisotropic flow. 
Redundancies found in the CMS data suggest that
these particle number fluctuations contribute significantly
to subleading components and may completely dominate
over the fluctuations of anisotropic flow that are 
nominally being measured.
The importance of multiplicity fluctuations to the standard PCA of flow fluctuations is a result of 
the remarkable sensitivity of the subleading PCA mode, the small size of the actual subleading flow $V_n^{(2)}$, and, of course,  
the choice of the covariance matrix of Eq.~(\ref{eq:flowsum}) in 
Refs.~\cite{Bhalerao:2014mua,Sirunyan:2017gyb}.  
Because particle number fluctuations can be measured separately and directly, they represent a redundant and unwanted background to principal component analyses of anisotropic flow.

 This led us to propose the PCA of $V_{n\Delta}$ and $V_{n\Delta}^R$, as defined in Eqs.~(\ref{eq:realdef-Vcov}) and (\ref{eq:pseudonorm-covmat}). 
The new observables are free of trivial contributions from multiplicity fluctuations, so that the new subleading PCA modes  
actually reveal fluctuations of anisotropies in the initial state. 
We tested the new proposed observables on simulated events generated by a hybrid hydrodynamic evolution model with event-by-event fluctuations. 
In Fig.~\ref{fig:PCA-R-hydro}, the new subleading modes appear to be relatively stable against changes in centrality, suggesting that they are not 
driven by the average geometry of the system. 
Also, the new proposed PCA subleading mode of anisotropic flow fluctuation $V_{n}^{R(2)}$, obtained from the covariance matrix $V_{n\Delta}^{R}$, is seen to reasonably 
reproduce $V_{n}^{(2)}$, obtained from the per particle flow covariance matrix $V_{n\Delta}$. 

The main advantage of the proposed observables over the factorization breaking measure $r_n(p_T^a,p_T^b)$ of Ref.~\cite{Gardim:2012im} 
is that they isolate linearly uncorrelated modes, making their physical content more transparent. 
They also allow for better, more compact, visualization, because the modes are functions of a single momentum variable. 
Furthermore, $r_n(p_T^a,p_T^b)$ only measures the relative importance of flow fluctuations, rendering them nearly imperceptible 
in noncentral collisions, where $V_n^{(1)}(\bm p)$ is larger \cite{Bhalerao:2014mua,Khachatryan:2015oea,CMS:2013bza,Acharya:2017ino}. 

The new PCA observables are sensitive to details of two-particle correlations and could provide new constraints to models 
of the  hydrodynamic expansion and of the initial state of heavy-ion collisions. 
Moreover, they could furnish much-needed insight into the physics of smaller collision systems, such as proton-nucleus, 
where fluctuations of the initial anisotropies should play a major role. 

We stress that the positivity of the PCA eigenvalues provides a highly nontrivial check of the hydrodynamic picture, as noted in 
Ref.~\cite{Bhalerao:2014mua}. 
While a covariance matrix is necessarily positive semidefinite, this is not strictly the case for the matrices defined in 
Eqs.~(\ref{eq:realdef-Vcov}), (\ref{eq:flowsum}), and (\ref{eq:pseudonorm-covmat}) outside a hydrodynamic picture. 
This is due to the absence of self-correlation terms ($a=b$) and would be aggravated 
by the implementation of a rapidity gap in an experimental analysis. 
Within the hydrodynamic picture, on the other hand, $V_n(\bm p)$ does not depend on individual particles  
and both $V_{n\Delta}$ and $V_{n\Delta}^N$ are covariance matrices, with positive or vanishing eigenvalues \cite{Bhalerao:2014mua}.  
Even in this picture, $V_{n\Delta}^R$ is not actually a covariance matrix, which could be seen as a disadvantage.

\section*{Acknowledgments}

We are thankful to D.~Teaney, A.~Mazeliauskas, S.~Mohapatra, P.~Bo\.zek and Marcus Bleicher for fruitful discussions. 
This research was funded by FAPESP Grants No.~2016/13803-2 (D.D.C.),
No.~2016/24029-6 (M.L.), No.~2017/05685-2 (all), No.~2018/01245-0 (T.N.dS.) and No.~2018/07833-1 (M.H.). 
D.D.C., M.L., J.N., and J.T. thank CNPq for financial support.
We also acknowledge computing time provided by the Research Computing
Support Group at Rice University through agreement with the University of São Paulo.

\appendix

\section{Model comparison to Data}

\begin{figure*}[b]
  \centering
   \includegraphics[width=\textwidth]{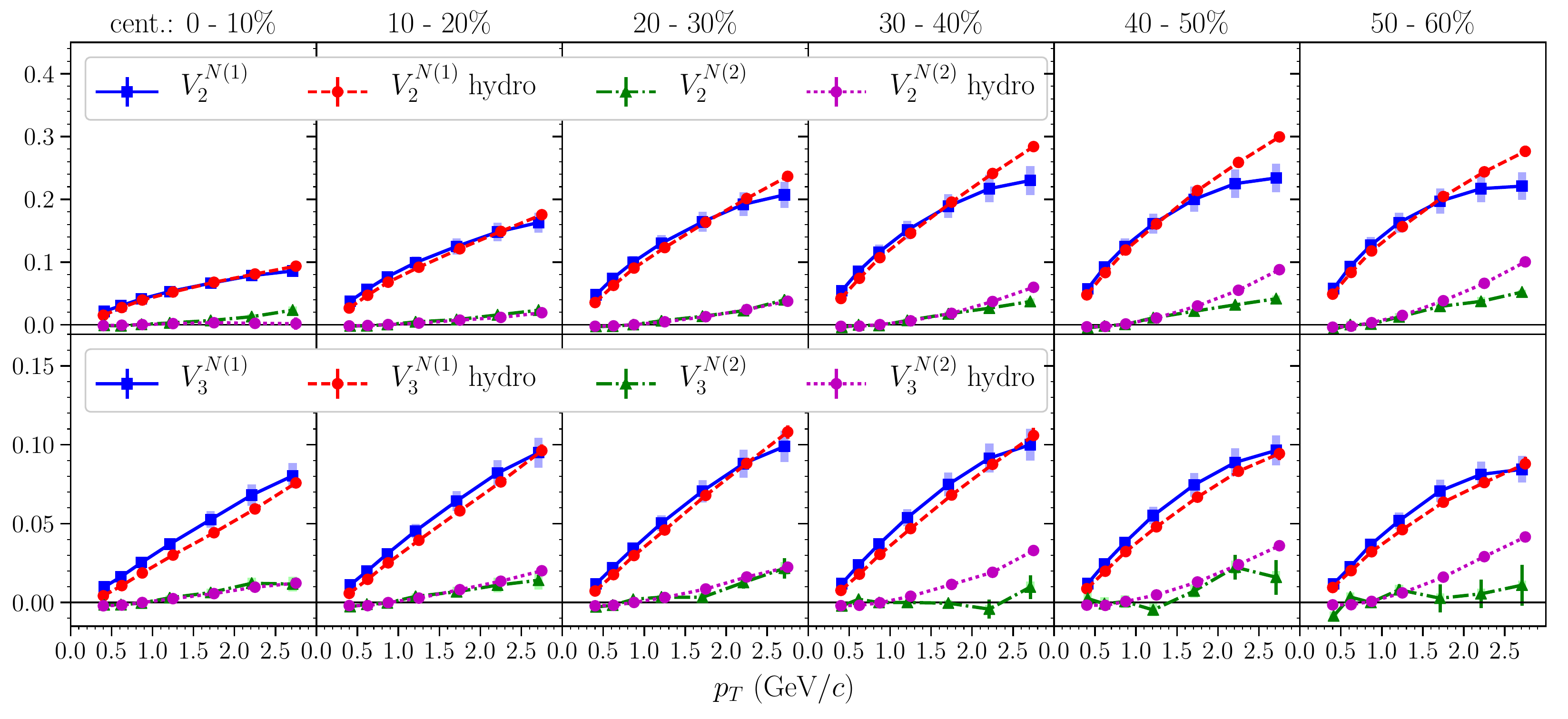}
   \caption{
   Comparison between the principal component analysis of anisotropic flow (as proposed in \cite{Bhalerao:2014mua}) 
   in both our hybrid hydrodynamic model ({\small TRENTO}+\small{MUSIC}+UrQMD) and published CMS Collaboration data \cite{Sirunyan:2017gyb}, 
   for     $Pb+Pb$ collisions at center-of-mass energy $\sqrt{s_{NN}} = 2.76$ TeV.
   }
   \label{fig:CMS-hydro}
 \end{figure*}
 
In this paper, we employ results from a hybrid hydrodynamic model, consisting of relativistic viscous hydrodynamics as implemented in {\small MUSIC} \cite{Schenke:2010nt,Schenke:2011bn} and  
 evolution of the hadron gas phase according to UrQMD \cite{Bass:1998ca,Bleicher:1999xi}. 
 Initial conditions were provided by {\small TRENTO} \cite{Moreland:2014oya} and 
parameter values were taken from the Bayesian analysis of Ref.~\cite{Bernhard:2018hnz}. 
Further details on this model can be found in Ref.~\cite{NunesdaSilva:2018viu}.

For completeness, we here include model results for the principal component analysis of event-by-event flow fluctuations, as originally proposed in Ref.~\cite{Bhalerao:2014mua}, 
for     $Pb+Pb$ collisions at center-of-mass energy $\sqrt{s_{NN}} = 2.76$ TeV.   
They are displayed in Fig.~\ref{fig:CMS-hydro}, where they are compared to published CMS Collaboration results \cite{Sirunyan:2017gyb} and a good agreement with data is verified. 

  \bibliography{multPCA}

\begin{thebibliography}{32}%
\makeatletter
\providecommand \@ifxundefined [1]{%
 \@ifx{#1\undefined}
}%
\providecommand \@ifnum [1]{%
 \ifnum #1\expandafter \@firstoftwo
 \else \expandafter \@secondoftwo
 \fi
}%
\providecommand \@ifx [1]{%
 \ifx #1\expandafter \@firstoftwo
 \else \expandafter \@secondoftwo
 \fi
}%
\providecommand \natexlab [1]{#1}%
\providecommand \enquote  [1]{``#1''}%
\providecommand \bibnamefont  [1]{#1}%
\providecommand \bibfnamefont [1]{#1}%
\providecommand \citenamefont [1]{#1}%
\providecommand \href@noop [0]{\@secondoftwo}%
\providecommand \href [0]{\begingroup \@sanitize@url \@href}%
\providecommand \@href[1]{\@@startlink{#1}\@@href}%
\providecommand \@@href[1]{\endgroup#1\@@endlink}%
\providecommand \@sanitize@url [0]{\catcode `\\12\catcode `\$12\catcode
  `\&12\catcode `\#12\catcode `\^12\catcode `\_12\catcode `\%12\relax}%
\providecommand \@@startlink[1]{}%
\providecommand \@@endlink[0]{}%
\providecommand \url  [0]{\begingroup\@sanitize@url \@url }%
\providecommand \@url [1]{\endgroup\@href {#1}{\urlprefix }}%
\providecommand \urlprefix  [0]{URL }%
\providecommand \Eprint [0]{\href }%
\providecommand \doibase [0]{http://dx.doi.org/}%
\providecommand \selectlanguage [0]{\@gobble}%
\providecommand \bibinfo  [0]{\@secondoftwo}%
\providecommand \bibfield  [0]{\@secondoftwo}%
\providecommand \translation [1]{[#1]}%
\providecommand \BibitemOpen [0]{}%
\providecommand \bibitemStop [0]{}%
\providecommand \bibitemNoStop [0]{.\EOS\space}%
\providecommand \EOS [0]{\spacefactor3000\relax}%
\providecommand \BibitemShut  [1]{\csname bibitem#1\endcsname}%
\let\auto@bib@innerbib\@empty
%</preamble>
\bibitem [{\citenamefont {Ollitrault}(1992)}]{Ollitrault:1992bk}%
  \BibitemOpen
  \bibfield  {author} {\bibinfo {author} {\bibfnamefont {Jean-Yves}\
  \bibnamefont {Ollitrault}},\ }\bibfield  {title} {\enquote {\bibinfo {title}
  {{Anisotropy as a signature of transverse collective flow}},}\ }\href
  {\doibase 10.1103/PhysRevD.46.229} {\bibfield  {journal} {\bibinfo  {journal}
  {Phys. Rev.}\ }\textbf {\bibinfo {volume} {D46}},\ \bibinfo {pages}
  {229--245} (\bibinfo {year} {1992})}\BibitemShut {NoStop}%
%%CITATION = PHRVA,D46,229;%%
\bibitem [{\citenamefont {Teaney}\ and\ \citenamefont
  {Yan}(2011)}]{Teaney:2010vd}%
  \BibitemOpen
  \bibfield  {author} {\bibinfo {author} {\bibfnamefont {Derek}\ \bibnamefont
  {Teaney}}\ and\ \bibinfo {author} {\bibfnamefont {Li}~\bibnamefont {Yan}},\
  }\bibfield  {title} {\enquote {\bibinfo {title} {{Triangularity and Dipole
  Asymmetry in Heavy Ion Collisions}},}\ }\href {\doibase
  10.1103/PhysRevC.83.064904} {\bibfield  {journal} {\bibinfo  {journal} {Phys.
  Rev.}\ }\textbf {\bibinfo {volume} {C83}},\ \bibinfo {pages} {064904}
  (\bibinfo {year} {2011})},\ \Eprint {http://arxiv.org/abs/1010.1876}
  {arXiv:1010.1876 [nucl-th]} \BibitemShut {NoStop}%
%%CITATION = ARXIV:1010.1876;%%
\bibitem [{\citenamefont {Gardim}\ \emph {et~al.}(2012)\citenamefont {Gardim},
  \citenamefont {Grassi}, \citenamefont {Luzum},\ and\ \citenamefont
  {Ollitrault}}]{Gardim:2011xv}%
  \BibitemOpen
  \bibfield  {author} {\bibinfo {author} {\bibfnamefont {Fernando~G.}\
  \bibnamefont {Gardim}}, \bibinfo {author} {\bibfnamefont {Frederique}\
  \bibnamefont {Grassi}}, \bibinfo {author} {\bibfnamefont {Matthew}\
  \bibnamefont {Luzum}}, \ and\ \bibinfo {author} {\bibfnamefont {Jean-Yves}\
  \bibnamefont {Ollitrault}},\ }\bibfield  {title} {\enquote {\bibinfo {title}
  {{Mapping the hydrodynamic response to the initial geometry in heavy-ion
  collisions}},}\ }\href {\doibase 10.1103/PhysRevC.85.024908} {\bibfield
  {journal} {\bibinfo  {journal} {Phys. Rev.}\ }\textbf {\bibinfo {volume}
  {C85}},\ \bibinfo {pages} {024908} (\bibinfo {year} {2012})},\ \Eprint
  {http://arxiv.org/abs/1111.6538} {arXiv:1111.6538 [nucl-th]} \BibitemShut
  {NoStop}%
%%CITATION = ARXIV:1111.6538;%%
\bibitem [{\citenamefont {Gardim}\ \emph {et~al.}(2018)\citenamefont {Gardim},
  \citenamefont {Grassi}, \citenamefont {Ishida}, \citenamefont {Luzum},
  \citenamefont {Magalhães},\ and\ \citenamefont
  {Noronha-Hostler}}]{Gardim:2017ruc}%
  \BibitemOpen
  \bibfield  {author} {\bibinfo {author} {\bibfnamefont {Fernando~G.}\
  \bibnamefont {Gardim}}, \bibinfo {author} {\bibfnamefont {Frédérique}\
  \bibnamefont {Grassi}}, \bibinfo {author} {\bibfnamefont {Pedro}\
  \bibnamefont {Ishida}}, \bibinfo {author} {\bibfnamefont {Matthew}\
  \bibnamefont {Luzum}}, \bibinfo {author} {\bibfnamefont {Pablo~S.}\
  \bibnamefont {Magalhães}}, \ and\ \bibinfo {author} {\bibfnamefont
  {Jacquelyn}\ \bibnamefont {Noronha-Hostler}},\ }\bibfield  {title} {\enquote
  {\bibinfo {title} {{Sensitivity of observables to coarse-graining size in
  heavy-ion collisions}},}\ }\href {\doibase 10.1103/PhysRevC.97.064919}
  {\bibfield  {journal} {\bibinfo  {journal} {Phys. Rev.}\ }\textbf {\bibinfo
  {volume} {C97}},\ \bibinfo {pages} {064919} (\bibinfo {year} {2018})},\
  \Eprint {http://arxiv.org/abs/1712.03912} {arXiv:1712.03912 [nucl-th]}
  \BibitemShut {NoStop}%
%%CITATION = ARXIV:1712.03912;%%
\bibitem [{\citenamefont {Kozlov}\ \emph
  {et~al.}(2014{\natexlab{a}})\citenamefont {Kozlov}, \citenamefont {Luzum},
  \citenamefont {Denicol}, \citenamefont {Jeon},\ and\ \citenamefont
  {Gale}}]{Kozlov:2014fqa}%
  \BibitemOpen
  \bibfield  {author} {\bibinfo {author} {\bibfnamefont {Igor}\ \bibnamefont
  {Kozlov}}, \bibinfo {author} {\bibfnamefont {Matthew}\ \bibnamefont {Luzum}},
  \bibinfo {author} {\bibfnamefont {Gabriel}\ \bibnamefont {Denicol}}, \bibinfo
  {author} {\bibfnamefont {Sangyong}\ \bibnamefont {Jeon}}, \ and\ \bibinfo
  {author} {\bibfnamefont {Charles}\ \bibnamefont {Gale}},\ }\bibfield  {title}
  {\enquote {\bibinfo {title} {{Transverse momentum structure of pair
  correlations as a signature of collective behavior in small collision
  systems}},}\ }\href@noop {} {\  (\bibinfo {year} {2014}{\natexlab{a}})},\
  \Eprint {http://arxiv.org/abs/1405.3976} {arXiv:1405.3976 [nucl-th]}
  \BibitemShut {NoStop}%
%%CITATION = ARXIV:1405.3976;%%
\bibitem [{\citenamefont {Noronha-Hostler}\ \emph {et~al.}(2016)\citenamefont
  {Noronha-Hostler}, \citenamefont {Noronha},\ and\ \citenamefont
  {Gyulassy}}]{Noronha-Hostler:2015coa}%
  \BibitemOpen
  \bibfield  {author} {\bibinfo {author} {\bibfnamefont {Jacquelyn}\
  \bibnamefont {Noronha-Hostler}}, \bibinfo {author} {\bibfnamefont {Jorge}\
  \bibnamefont {Noronha}}, \ and\ \bibinfo {author} {\bibfnamefont {Miklos}\
  \bibnamefont {Gyulassy}},\ }\bibfield  {title} {\enquote {\bibinfo {title}
  {{Sensitivity of flow harmonics to subnucleon scale fluctuations in heavy ion
  collisions}},}\ }\href {\doibase 10.1103/PhysRevC.93.024909} {\bibfield
  {journal} {\bibinfo  {journal} {Phys. Rev.}\ }\textbf {\bibinfo {volume}
  {C93}},\ \bibinfo {pages} {024909} (\bibinfo {year} {2016})},\ \Eprint
  {http://arxiv.org/abs/1508.02455} {arXiv:1508.02455 [nucl-th]} \BibitemShut
  {NoStop}%
%%CITATION = ARXIV:1508.02455;%%
\bibitem [{\citenamefont {Gardim}\ \emph {et~al.}(2013)\citenamefont {Gardim},
  \citenamefont {Grassi}, \citenamefont {Luzum},\ and\ \citenamefont
  {Ollitrault}}]{Gardim:2012im}%
  \BibitemOpen
  \bibfield  {author} {\bibinfo {author} {\bibfnamefont {Fernando~G.}\
  \bibnamefont {Gardim}}, \bibinfo {author} {\bibfnamefont {Frederique}\
  \bibnamefont {Grassi}}, \bibinfo {author} {\bibfnamefont {Matthew}\
  \bibnamefont {Luzum}}, \ and\ \bibinfo {author} {\bibfnamefont {Jean-Yves}\
  \bibnamefont {Ollitrault}},\ }\bibfield  {title} {\enquote {\bibinfo {title}
  {{Breaking of factorization of two-particle correlations in
  hydrodynamics}},}\ }\href {\doibase 10.1103/PhysRevC.87.031901} {\bibfield
  {journal} {\bibinfo  {journal} {Phys. Rev.}\ }\textbf {\bibinfo {volume}
  {C87}},\ \bibinfo {pages} {031901(R)} (\bibinfo {year} {2013})},\ \Eprint
  {http://arxiv.org/abs/1211.0989} {arXiv:1211.0989 [nucl-th]} \BibitemShut
  {NoStop}%
%%CITATION = ARXIV:1211.0989;%%
\bibitem [{\citenamefont {Heinz}\ \emph {et~al.}(2013)\citenamefont {Heinz},
  \citenamefont {Qiu},\ and\ \citenamefont {Shen}}]{Heinz:2013bua}%
  \BibitemOpen
  \bibfield  {author} {\bibinfo {author} {\bibfnamefont {Ulrich}\ \bibnamefont
  {Heinz}}, \bibinfo {author} {\bibfnamefont {Zhi}\ \bibnamefont {Qiu}}, \ and\
  \bibinfo {author} {\bibfnamefont {Chun}\ \bibnamefont {Shen}},\ }\bibfield
  {title} {\enquote {\bibinfo {title} {{Fluctuating flow angles and anisotropic
  flow measurements}},}\ }\href {\doibase 10.1103/PhysRevC.87.034913}
  {\bibfield  {journal} {\bibinfo  {journal} {Phys. Rev.}\ }\textbf {\bibinfo
  {volume} {C87}},\ \bibinfo {pages} {034913} (\bibinfo {year} {2013})},\
  \Eprint {http://arxiv.org/abs/1302.3535} {arXiv:1302.3535 [nucl-th]}
  \BibitemShut {NoStop}%
%%CITATION = ARXIV:1302.3535;%%
\bibitem [{\citenamefont {Kozlov}\ \emph
  {et~al.}(2014{\natexlab{b}})\citenamefont {Kozlov}, \citenamefont {Luzum},
  \citenamefont {Denicol}, \citenamefont {Jeon},\ and\ \citenamefont
  {Gale}}]{Kozlov:2014hya}%
  \BibitemOpen
  \bibfield  {author} {\bibinfo {author} {\bibfnamefont {I.}~\bibnamefont
  {Kozlov}}, \bibinfo {author} {\bibfnamefont {Matthew}\ \bibnamefont {Luzum}},
  \bibinfo {author} {\bibfnamefont {Gabriel~S.}\ \bibnamefont {Denicol}},
  \bibinfo {author} {\bibfnamefont {Sangyong}\ \bibnamefont {Jeon}}, \ and\
  \bibinfo {author} {\bibfnamefont {Charles}\ \bibnamefont {Gale}},\ }\bibfield
   {title} {\enquote {\bibinfo {title} {{Signatures of collective behavior in
  small systems}},}\ }\bibfield  {booktitle} {\emph {\bibinfo {booktitle} {{in
  Proceedings of the 24th International Conference on Ultra-Relativistic
  Nucleus-Nucleus Collisions (Quark Matter 2014): Darmstadt, Germany, May
  19-24, 2014}}},\ }\href {\doibase 10.1016/j.nuclphysa.2014.09.054} {\bibfield
   {journal} {\bibinfo  {journal} {Nucl. Phys.}\ }\textbf {\bibinfo {volume}
  {A931}},\ \bibinfo {pages} {1045--1050} (\bibinfo {year}
  {2014}{\natexlab{b}})},\ \Eprint {http://arxiv.org/abs/1412.3147}
  {arXiv:1412.3147 [nucl-th]} \BibitemShut {NoStop}%
%%CITATION = ARXIV:1412.3147;%%
\bibitem [{\citenamefont {Shen}\ \emph {et~al.}(2015)\citenamefont {Shen},
  \citenamefont {Qiu},\ and\ \citenamefont {Heinz}}]{Shen:2015qta}%
  \BibitemOpen
  \bibfield  {author} {\bibinfo {author} {\bibfnamefont {Chun}\ \bibnamefont
  {Shen}}, \bibinfo {author} {\bibfnamefont {Zhi}\ \bibnamefont {Qiu}}, \ and\
  \bibinfo {author} {\bibfnamefont {Ulrich}\ \bibnamefont {Heinz}},\ }\bibfield
   {title} {\enquote {\bibinfo {title} {{Shape and flow fluctuations in
  ultracentral Pb + Pb collisions at the energies available at the CERN Large
  Hadron Collider}},}\ }\href {\doibase 10.1103/PhysRevC.92.014901} {\bibfield
  {journal} {\bibinfo  {journal} {Phys. Rev.}\ }\textbf {\bibinfo {volume}
  {C92}},\ \bibinfo {pages} {014901} (\bibinfo {year} {2015})},\ \Eprint
  {http://arxiv.org/abs/1502.04636} {arXiv:1502.04636 [nucl-th]} \BibitemShut
  {NoStop}%
%%CITATION = ARXIV:1502.04636;%%
\bibitem [{\citenamefont {Zhao}\ \emph {et~al.}(2017)\citenamefont {Zhao},
  \citenamefont {Xu},\ and\ \citenamefont {Song}}]{Zhao:2017yhj}%
  \BibitemOpen
  \bibfield  {author} {\bibinfo {author} {\bibfnamefont {Wenbin}\ \bibnamefont
  {Zhao}}, \bibinfo {author} {\bibfnamefont {Hao-jie}\ \bibnamefont {Xu}}, \
  and\ \bibinfo {author} {\bibfnamefont {Huichao}\ \bibnamefont {Song}},\
  }\bibfield  {title} {\enquote {\bibinfo {title} {{Collective flow in 2.76 A
  TeV and 5.02 A TeV Pb+Pb collisions}},}\ }\href {\doibase
  10.1140/epjc/s10052-017-5186-x} {\bibfield  {journal} {\bibinfo  {journal}
  {Eur. Phys. J.}\ }\textbf {\bibinfo {volume} {C77}},\ \bibinfo {pages} {645}
  (\bibinfo {year} {2017})},\ \Eprint {http://arxiv.org/abs/1703.10792}
  {arXiv:1703.10792 [nucl-th]} \BibitemShut {NoStop}%
%%CITATION = ARXIV:1703.10792;%%
\bibitem [{\citenamefont {Bożek}(2018)}]{Bozek:2018nne}%
  \BibitemOpen
  \bibfield  {author} {\bibinfo {author} {\bibfnamefont {Piotr}\ \bibnamefont
  {Bożek}},\ }\bibfield  {title} {\enquote {\bibinfo {title} {{Angle and
  magnitude decorrelation in the factorization breaking of collective flow}},}\
  }\href {\doibase 10.1103/PhysRevC.98.064906} {\bibfield  {journal} {\bibinfo
  {journal} {Phys. Rev.}\ }\textbf {\bibinfo {volume} {C98}},\ \bibinfo {pages}
  {064906} (\bibinfo {year} {2018})},\ \Eprint
  {http://arxiv.org/abs/1808.04248} {arXiv:1808.04248 [nucl-th]} \BibitemShut
  {NoStop}%
%%CITATION = ARXIV:1808.04248;%%
\bibitem [{\citenamefont {Bhalerao}\ \emph {et~al.}(2015)\citenamefont
  {Bhalerao}, \citenamefont {Ollitrault}, \citenamefont {Pal},\ and\
  \citenamefont {Teaney}}]{Bhalerao:2014mua}%
  \BibitemOpen
  \bibfield  {author} {\bibinfo {author} {\bibfnamefont {Rajeev~S.}\
  \bibnamefont {Bhalerao}}, \bibinfo {author} {\bibfnamefont {Jean-Yves}\
  \bibnamefont {Ollitrault}}, \bibinfo {author} {\bibfnamefont {Subrata}\
  \bibnamefont {Pal}}, \ and\ \bibinfo {author} {\bibfnamefont {Derek}\
  \bibnamefont {Teaney}},\ }\bibfield  {title} {\enquote {\bibinfo {title}
  {{Principal component analysis of event-by-event fluctuations}},}\ }\href
  {\doibase 10.1103/PhysRevLett.114.152301} {\bibfield  {journal} {\bibinfo
  {journal} {Phys. Rev. Lett.}\ }\textbf {\bibinfo {volume} {114}},\ \bibinfo
  {pages} {152301} (\bibinfo {year} {2015})},\ \Eprint
  {http://arxiv.org/abs/1410.7739} {arXiv:1410.7739 [nucl-th]} \BibitemShut
  {NoStop}%
%%CITATION = ARXIV:1410.7739;%%
\bibitem [{\citenamefont {Mazeliauskas}\ and\ \citenamefont
  {Teaney}(2015)}]{Mazeliauskas:2015vea}%
  \BibitemOpen
  \bibfield  {author} {\bibinfo {author} {\bibfnamefont {Aleksas}\ \bibnamefont
  {Mazeliauskas}}\ and\ \bibinfo {author} {\bibfnamefont {Derek}\ \bibnamefont
  {Teaney}},\ }\bibfield  {title} {\enquote {\bibinfo {title} {{Subleading
  harmonic flows in hydrodynamic simulations of heavy ion collisions}},}\
  }\href {\doibase 10.1103/PhysRevC.91.044902} {\bibfield  {journal} {\bibinfo
  {journal} {Phys. Rev.}\ }\textbf {\bibinfo {volume} {C91}},\ \bibinfo {pages}
  {044902} (\bibinfo {year} {2015})},\ \Eprint
  {http://arxiv.org/abs/1501.03138} {arXiv:1501.03138 [nucl-th]} \BibitemShut
  {NoStop}%
%%CITATION = ARXIV:1501.03138;%%
\bibitem [{\citenamefont {Mazeliauskas}\ and\ \citenamefont
  {Teaney}(2016)}]{Mazeliauskas:2015efa}%
  \BibitemOpen
  \bibfield  {author} {\bibinfo {author} {\bibfnamefont {Aleksas}\ \bibnamefont
  {Mazeliauskas}}\ and\ \bibinfo {author} {\bibfnamefont {Derek}\ \bibnamefont
  {Teaney}},\ }\bibfield  {title} {\enquote {\bibinfo {title} {{Fluctuations of
  harmonic and radial flow in heavy ion collisions with principal
  components}},}\ }\href {\doibase 10.1103/PhysRevC.93.024913} {\bibfield
  {journal} {\bibinfo  {journal} {Phys. Rev.}\ }\textbf {\bibinfo {volume}
  {C93}},\ \bibinfo {pages} {024913} (\bibinfo {year} {2016})},\ \Eprint
  {http://arxiv.org/abs/1509.07492} {arXiv:1509.07492 [nucl-th]} \BibitemShut
  {NoStop}%
%%CITATION = ARXIV:1509.07492;%%
\bibitem [{\citenamefont {Cirkovic}\ \emph {et~al.}(2017)\citenamefont
  {Cirkovic}, \citenamefont {Devetak}, \citenamefont {Dordevic}, \citenamefont
  {Milosevic},\ and\ \citenamefont {Stojanovic}}]{Cirkovic:2016kxt}%
  \BibitemOpen
  \bibfield  {author} {\bibinfo {author} {\bibfnamefont {P.}~\bibnamefont
  {Cirkovic}}, \bibinfo {author} {\bibfnamefont {D.}~\bibnamefont {Devetak}},
  \bibinfo {author} {\bibfnamefont {M.}~\bibnamefont {Dordevic}}, \bibinfo
  {author} {\bibfnamefont {J.}~\bibnamefont {Milosevic}}, \ and\ \bibinfo
  {author} {\bibfnamefont {M.}~\bibnamefont {Stojanovic}},\ }\bibfield  {title}
  {\enquote {\bibinfo {title} {{Sub-leading flow modes in PbPb collisions at
  $\sqrt{s_{NN}}$ = 2.76 TeV from HYDJET++ model}},}\ }\href {\doibase
  10.1088/1674-1137/41/7/074001} {\bibfield  {journal} {\bibinfo  {journal}
  {Chin. Phys.}\ }\textbf {\bibinfo {volume} {C41}},\ \bibinfo {pages} {074001}
  (\bibinfo {year} {2017})},\ \Eprint {http://arxiv.org/abs/1611.06602}
  {arXiv:1611.06602 [nucl-ex]} \BibitemShut {NoStop}%
%%CITATION = ARXIV:1611.06602;%%
\bibitem [{\citenamefont {Bozek}(2018)}]{Bozek:2017thv}%
  \BibitemOpen
  \bibfield  {author} {\bibinfo {author} {\bibfnamefont {Piotr}\ \bibnamefont
  {Bozek}},\ }\bibfield  {title} {\enquote {\bibinfo {title} {{Principal
  component analysis of the nonlinear coupling of harmonic modes in heavy-ion
  collisions}},}\ }\href {\doibase 10.1103/PhysRevC.97.034905} {\bibfield
  {journal} {\bibinfo  {journal} {Phys. Rev.}\ }\textbf {\bibinfo {volume}
  {C97}},\ \bibinfo {pages} {034905} (\bibinfo {year} {2018})},\ \Eprint
  {http://arxiv.org/abs/1711.07773} {arXiv:1711.07773 [nucl-th]} \BibitemShut
  {NoStop}%
%%CITATION = ARXIV:1711.07773;%%
\bibitem [{\citenamefont {Sirunyan}\ \emph {et~al.}(2017)\citenamefont
  {Sirunyan} \emph {et~al.}}]{Sirunyan:2017gyb}%
  \BibitemOpen
  \bibfield  {author} {\bibinfo {author} {\bibfnamefont {A.~M.}\ \bibnamefont
  {Sirunyan}} \emph {et~al.} (\bibinfo {collaboration} {CMS Collaboration}),\
  }\bibfield  {title} {\enquote {\bibinfo {title} {{Principal-component
  analysis of two-particle azimuthal correlations in PbPb and $p$Pb collisions
  at CMS}},}\ }\href {\doibase 10.1103/PhysRevC.96.064902} {\bibfield
  {journal} {\bibinfo  {journal} {Phys. Rev.}\ }\textbf {\bibinfo {volume}
  {C96}},\ \bibinfo {pages} {064902} (\bibinfo {year} {2017})},\ \Eprint
  {http://arxiv.org/abs/1708.07113} {arXiv:1708.07113 [nucl-ex]} \BibitemShut
  {NoStop}%
%%CITATION = ARXIV:1708.07113;%%
\bibitem [{\citenamefont {Khachatryan}\ \emph {et~al.}(2015)\citenamefont
  {Khachatryan} \emph {et~al.}}]{Khachatryan:2015oea}%
  \BibitemOpen
  \bibfield  {author} {\bibinfo {author} {\bibfnamefont {Vardan}\ \bibnamefont
  {Khachatryan}} \emph {et~al.} (\bibinfo {collaboration} {CMS
  Collaboration}),\ }\bibfield  {title} {\enquote {\bibinfo {title} {{Evidence
  for transverse-momentum- and pseudorapidity-dependent event-plane
  fluctuations in PbPb and $p$Pb collisions}},}\ }\href {\doibase
  10.1103/PhysRevC.92.034911} {\bibfield  {journal} {\bibinfo  {journal} {Phys.
  Rev.}\ }\textbf {\bibinfo {volume} {C92}},\ \bibinfo {pages} {034911}
  (\bibinfo {year} {2015})},\ \Eprint {http://arxiv.org/abs/1503.01692}
  {arXiv:1503.01692 [nucl-ex]} \BibitemShut {NoStop}%
%%CITATION = ARXIV:1503.01692;%%
\bibitem [{\citenamefont {Chatrchyan}\ \emph {et~al.}(2014)\citenamefont
  {Chatrchyan} \emph {et~al.}}]{CMS:2013bza}%
  \BibitemOpen
  \bibfield  {author} {\bibinfo {author} {\bibfnamefont {Serguei}\ \bibnamefont
  {Chatrchyan}} \emph {et~al.} (\bibinfo {collaboration} {CMS}),\ }\bibfield
  {title} {\enquote {\bibinfo {title} {{Studies of azimuthal dihadron
  correlations in ultra-central PbPb collisions at $\sqrt{s_{NN}} =$ 2.76
  TeV}},}\ }\href {\doibase 10.1007/JHEP02(2014)088} {\bibfield  {journal}
  {\bibinfo  {journal} {JHEP}\ }\textbf {\bibinfo {volume} {02}},\ \bibinfo
  {pages} {088} (\bibinfo {year} {2014})},\ \Eprint
  {http://arxiv.org/abs/1312.1845} {arXiv:1312.1845 [nucl-ex]} \BibitemShut
  {NoStop}%
%%CITATION = ARXIV:1312.1845;%%
\bibitem [{\citenamefont {Acharya}\ \emph {et~al.}(2017)\citenamefont {Acharya}
  \emph {et~al.}}]{Acharya:2017ino}%
  \BibitemOpen
  \bibfield  {author} {\bibinfo {author} {\bibfnamefont {Shreyasi}\
  \bibnamefont {Acharya}} \emph {et~al.} (\bibinfo {collaboration} {ALICE
  Collaboration}),\ }\bibfield  {title} {\enquote {\bibinfo {title} {{Searches
  for transverse momentum dependent flow vector fluctuations in Pb-Pb and p-Pb
  collisions at the LHC}},}\ }\href {\doibase 10.1007/JHEP09(2017)032}
  {\bibfield  {journal} {\bibinfo  {journal} {JHEP}\ }\textbf {\bibinfo
  {volume} {09}},\ \bibinfo {pages} {032} (\bibinfo {year} {2017})},\ \Eprint
  {http://arxiv.org/abs/1707.05690} {arXiv:1707.05690 [nucl-ex]} \BibitemShut
  {NoStop}%
%%CITATION = ARXIV:1707.05690;%%
\bibitem [{\citenamefont {Jolliffe}(2005)}]{doi:10.1002/0470013192.bsa501}%
  \BibitemOpen
  \bibfield  {author} {\bibinfo {author} {\bibfnamefont {Ian}\ \bibnamefont
  {Jolliffe}},\ }\enquote {\bibinfo {title} {Principal component analysis},}\
  in\ \href {\doibase 10.1002/0470013192.bsa501} {\emph {\bibinfo {booktitle}
  {Encyclopedia of Statistics in Behavioral Science}}}\ (\bibinfo  {publisher}
  {Wiley Online Library},\ \bibinfo {year} {2005})\BibitemShut {NoStop}%
\bibitem [{\citenamefont {Gardim}\ \emph {et~al.}(2019)\citenamefont {Gardim},
  \citenamefont {Grassi}, \citenamefont {Ishida}, \citenamefont {Luzum},\ and\
  \citenamefont {Ollitrault}}]{Gardim:2019iah}%
  \BibitemOpen
  \bibfield  {author} {\bibinfo {author} {\bibfnamefont {Fernando~G.}\
  \bibnamefont {Gardim}}, \bibinfo {author} {\bibfnamefont {Frederique}\
  \bibnamefont {Grassi}}, \bibinfo {author} {\bibfnamefont {Pedro}\
  \bibnamefont {Ishida}}, \bibinfo {author} {\bibfnamefont {Matthew}\
  \bibnamefont {Luzum}}, \ and\ \bibinfo {author} {\bibfnamefont {Jean-Yves}\
  \bibnamefont {Ollitrault}},\ }\bibfield  {title} {\enquote {\bibinfo {title}
  {{$p_T$-Dependent Particle Number Fluctuations From Principal Component
  Analyses in Hydrodynamic Simulations of Heavy-Ion Collisions}},}\ }\href@noop
  {} {\  (\bibinfo {year} {2019})},\ \Eprint {http://arxiv.org/abs/1906.03045}
  {arXiv:1906.03045 [nucl-th]} \BibitemShut {NoStop}%
%%CITATION = ARXIV:1906.03045;%%
\bibitem [{\citenamefont {Heinz}\ and\ \citenamefont
  {Snellings}(2013)}]{Heinz:2013th}%
  \BibitemOpen
  \bibfield  {author} {\bibinfo {author} {\bibfnamefont {Ulrich}\ \bibnamefont
  {Heinz}}\ and\ \bibinfo {author} {\bibfnamefont {Raimond}\ \bibnamefont
  {Snellings}},\ }\bibfield  {title} {\enquote {\bibinfo {title} {{Collective
  flow and viscosity in relativistic heavy-ion collisions}},}\ }\href {\doibase
  10.1146/annurev-nucl-102212-170540} {\bibfield  {journal} {\bibinfo
  {journal} {Ann. Rev. Nucl. Part. Sci.}\ }\textbf {\bibinfo {volume} {63}},\
  \bibinfo {pages} {123--151} (\bibinfo {year} {2013})},\ \Eprint
  {http://arxiv.org/abs/1301.2826} {arXiv:1301.2826 [nucl-th]} \BibitemShut
  {NoStop}%
%%CITATION = ARXIV:1301.2826;%%
\bibitem [{\citenamefont {Nunes~da Silva}\ \emph {et~al.}(2019)\citenamefont
  {Nunes~da Silva}, \citenamefont {Dobrigkeit~Chinellato}, \citenamefont
  {Derradi De~Souza}, \citenamefont {Hippert}, \citenamefont {Luzum},
  \citenamefont {Noronha},\ and\ \citenamefont
  {Takahashi}}]{NunesdaSilva:2018viu}%
  \BibitemOpen
  \bibfield  {author} {\bibinfo {author} {\bibfnamefont {Tiago}\ \bibnamefont
  {Nunes~da Silva}}, \bibinfo {author} {\bibfnamefont {David}\ \bibnamefont
  {Dobrigkeit~Chinellato}}, \bibinfo {author} {\bibfnamefont {Rafael}\
  \bibnamefont {Derradi De~Souza}}, \bibinfo {author} {\bibfnamefont
  {Maurício}\ \bibnamefont {Hippert}}, \bibinfo {author} {\bibfnamefont
  {Matthew}\ \bibnamefont {Luzum}}, \bibinfo {author} {\bibfnamefont {Jorge}\
  \bibnamefont {Noronha}}, \ and\ \bibinfo {author} {\bibfnamefont {Jun}\
  \bibnamefont {Takahashi}},\ }\bibfield  {title} {\enquote {\bibinfo {title}
  {{Testing a best-fit hydrodynamical model using PCA}},}\ }\bibfield
  {booktitle} {\emph {\bibinfo {booktitle} {{in Proceedings of Hot Quarks 2018:
  Workshop for Young Scientists on the Physics of Ultrarelativistic
  Nucleus-Nucleus Collisions (HQ2018): De Krim, Texel Island, Netherlands,
  September 7-14, 2018}}},\ }\href {\doibase 10.3390/proceedings2019010005}
  {\bibfield  {journal} {\bibinfo  {journal} {MDPI Proc.}\ }\textbf {\bibinfo
  {volume} {10}},\ \bibinfo {pages} {5} (\bibinfo {year} {2019})},\ \Eprint
  {http://arxiv.org/abs/1811.05048} {arXiv:1811.05048 [nucl-th]} \BibitemShut
  {NoStop}%
%%CITATION = ARXIV:1811.05048;%%
\bibitem [{\citenamefont {Schenke}\ \emph {et~al.}(2010)\citenamefont
  {Schenke}, \citenamefont {Jeon},\ and\ \citenamefont
  {Gale}}]{Schenke:2010nt}%
  \BibitemOpen
  \bibfield  {author} {\bibinfo {author} {\bibfnamefont {Bjoern}\ \bibnamefont
  {Schenke}}, \bibinfo {author} {\bibfnamefont {Sangyong}\ \bibnamefont
  {Jeon}}, \ and\ \bibinfo {author} {\bibfnamefont {Charles}\ \bibnamefont
  {Gale}},\ }\bibfield  {title} {\enquote {\bibinfo {title} {{(3+1)D
  hydrodynamic simulation of relativistic heavy-ion collisions}},}\ }\href
  {\doibase 10.1103/PhysRevC.82.014903} {\bibfield  {journal} {\bibinfo
  {journal} {Phys. Rev.}\ }\textbf {\bibinfo {volume} {C82}},\ \bibinfo {pages}
  {014903} (\bibinfo {year} {2010})},\ \Eprint {http://arxiv.org/abs/1004.1408}
  {arXiv:1004.1408 [hep-ph]} \BibitemShut {NoStop}%
%%CITATION = ARXIV:1004.1408;%%
\bibitem [{\citenamefont {Schenke}\ \emph {et~al.}(2012)\citenamefont
  {Schenke}, \citenamefont {Jeon},\ and\ \citenamefont
  {Gale}}]{Schenke:2011bn}%
  \BibitemOpen
  \bibfield  {author} {\bibinfo {author} {\bibfnamefont {Bjorn}\ \bibnamefont
  {Schenke}}, \bibinfo {author} {\bibfnamefont {Sangyong}\ \bibnamefont
  {Jeon}}, \ and\ \bibinfo {author} {\bibfnamefont {Charles}\ \bibnamefont
  {Gale}},\ }\bibfield  {title} {\enquote {\bibinfo {title} {{Higher flow
  harmonics from (3+1)D event-by-event viscous hydrodynamics}},}\ }\href
  {\doibase 10.1103/PhysRevC.85.024901} {\bibfield  {journal} {\bibinfo
  {journal} {Phys. Rev.}\ }\textbf {\bibinfo {volume} {C85}},\ \bibinfo {pages}
  {024901} (\bibinfo {year} {2012})},\ \Eprint {http://arxiv.org/abs/1109.6289}
  {arXiv:1109.6289 [hep-ph]} \BibitemShut {NoStop}%
%%CITATION = ARXIV:1109.6289;%%
\bibitem [{\citenamefont {Bass}\ \emph {et~al.}(1998)\citenamefont {Bass} \emph
  {et~al.}}]{Bass:1998ca}%
  \BibitemOpen
  \bibfield  {author} {\bibinfo {author} {\bibfnamefont {S.~A.}\ \bibnamefont
  {Bass}} \emph {et~al.},\ }\bibfield  {title} {\enquote {\bibinfo {title}
  {{Microscopic models for ultrarelativistic heavy ion collisions}},}\ }\href
  {\doibase 10.1016/S0146-6410(98)00058-1} {\bibfield  {journal} {\bibinfo
  {journal} {Prog. Part. Nucl. Phys.}\ }\textbf {\bibinfo {volume} {41}},\
  \bibinfo {pages} {255--369} (\bibinfo {year} {1998})},\ \bibinfo {note}
  {[Prog. Part. Nucl. Phys.41,225(1998)]},\ \Eprint
  {http://arxiv.org/abs/nucl-th/9803035} {arXiv:nucl-th/9803035 [nucl-th]}
  \BibitemShut {NoStop}%
%%CITATION = NUCL-TH/9803035;%%
\bibitem [{\citenamefont {Bleicher}\ \emph {et~al.}(1999)\citenamefont
  {Bleicher} \emph {et~al.}}]{Bleicher:1999xi}%
  \BibitemOpen
  \bibfield  {author} {\bibinfo {author} {\bibfnamefont {M.}~\bibnamefont
  {Bleicher}} \emph {et~al.},\ }\bibfield  {title} {\enquote {\bibinfo {title}
  {{Relativistic hadron hadron collisions in the ultrarelativistic quantum
  molecular dynamics model}},}\ }\href {\doibase 10.1088/0954-3899/25/9/308}
  {\bibfield  {journal} {\bibinfo  {journal} {J. Phys.}\ }\textbf {\bibinfo
  {volume} {G25}},\ \bibinfo {pages} {1859--1896} (\bibinfo {year} {1999})},\
  \Eprint {http://arxiv.org/abs/hep-ph/9909407} {arXiv:hep-ph/9909407 [hep-ph]}
  \BibitemShut {NoStop}%
%%CITATION = HEP-PH/9909407;%%
\bibitem [{\citenamefont {Moreland}\ \emph {et~al.}(2015)\citenamefont
  {Moreland}, \citenamefont {Bernhard},\ and\ \citenamefont
  {Bass}}]{Moreland:2014oya}%
  \BibitemOpen
  \bibfield  {author} {\bibinfo {author} {\bibfnamefont {J.~Scott}\
  \bibnamefont {Moreland}}, \bibinfo {author} {\bibfnamefont {Jonah~E.}\
  \bibnamefont {Bernhard}}, \ and\ \bibinfo {author} {\bibfnamefont
  {Steffen~A.}\ \bibnamefont {Bass}},\ }\bibfield  {title} {\enquote {\bibinfo
  {title} {{Alternative ansatz to wounded nucleon and binary collision scaling
  in high-energy nuclear collisions}},}\ }\href {\doibase
  10.1103/PhysRevC.92.011901} {\bibfield  {journal} {\bibinfo  {journal} {Phys.
  Rev.}\ }\textbf {\bibinfo {volume} {C92}},\ \bibinfo {pages} {011901(R)}
  (\bibinfo {year} {2015})},\ \Eprint {http://arxiv.org/abs/1412.4708}
  {arXiv:1412.4708 [nucl-th]} \BibitemShut {NoStop}%
%%CITATION = ARXIV:1412.4708;%%
\bibitem [{\citenamefont {Bernhard}(2018-04-19)}]{Bernhard:2018hnz}%
  \BibitemOpen
  \bibfield  {author} {\bibinfo {author} {\bibfnamefont {Jonah~E.}\
  \bibnamefont {Bernhard}},\ }\emph {\bibinfo {title} {{Bayesian parameter
  estimation for relativistic heavy-ion collisions}}},\ \href@noop {} {Ph.D.
  thesis},\ \bibinfo  {school} {Duke University} (\bibinfo {year}
  {2018-04-19}),\ \Eprint {http://arxiv.org/abs/1804.06469} {arXiv:1804.06469
  [nucl-th]} \BibitemShut {NoStop}%
%%CITATION = ARXIV:1804.06469;%%
\bibitem [{\citenamefont {Hippert}\ \emph {et~al.}()\citenamefont {Hippert},
  \citenamefont {Dobrigkeit~Chinellato}, \citenamefont {Luzum}, \citenamefont
  {Noronha}, \citenamefont {Nunes~da Silva},\ and\ \citenamefont
  {Takahashi}}]{inprogress}%
  \BibitemOpen
  \bibfield  {author} {\bibinfo {author} {\bibfnamefont {Mauricio}\
  \bibnamefont {Hippert}}, \bibinfo {author} {\bibfnamefont {David}\
  \bibnamefont {Dobrigkeit~Chinellato}}, \bibinfo {author} {\bibfnamefont
  {Matthew}\ \bibnamefont {Luzum}}, \bibinfo {author} {\bibfnamefont {Jorge}\
  \bibnamefont {Noronha}}, \bibinfo {author} {\bibfnamefont {Tiago}\
  \bibnamefont {Nunes~da Silva}}, \ and\ \bibinfo {author} {\bibfnamefont
  {Jun}\ \bibnamefont {Takahashi}},\ }\href@noop {} {\bibinfo  {journal}
  {(unpublished)}\ }\BibitemShut {NoStop}%
\end{thebibliography}%

\end{document}